\documentclass[fleqn,twoside]{article}
\usepackage{espcrc2}
\usepackage{epsfig}
\usepackage{amssymb}
\usepackage{graphicx}% Include figure files
\usepackage{dcolumn}% Align table columns on decimal point
\usepackage{bm}% bold math

\newcommand{\ttbs}{\char'134}
\newcommand{\AmS}{{\protect\the\textfont2
  A\kern-.1667em\lower.5ex\hbox{M}\kern-.125emS}}

\title{Tunable ratchet effects for vortices pinned by periodic magnetic dipole arrays
\tt\ttbs\thanks {Research supported in part by the Brazilian agencies CNPq, CAPES, FAPERJ,  and FUJB.}}

\author{Gilson Carneiro
\address{Instituto de F\'{\i}sica, Universidade Federal do Rio de Janeiro,  
C.P. 68528, 21941-972, Rio de Janeiro-RJ, Brasil }}

\begin{document}

\begin{abstract}

The ratchet effect is demonstrated theoretically for 
the simple model of a vortex in a thin superconducting film interacting with a periodic  array of magnetic dipoles placed in the vicinity of the film . The pinning potential for the vortex  is calculated in the London limit and found to break spatial inversion symmetry and to depend on the orientation of the magnetic dipole moments. The motion of the vortex at zero temperature driven by a force oscillating periodically in time  is investigated numerically.  Drift vortex motion consisting of  displacements by a translation vector of the dipole array during each period of oscillation is obtained and studied in detail. The direction of drift   
differs in general from that of the driving force, except if the driving force oscillates in a direction of high symmetry of the dipole array.  The vortex drift velocity  depends on the orientation of the magnetic moments, and can be  tuned by rotating the dipoles. It is pointed out that if the magnetic moments are free to rotate, the ratchet effect can be produced and tuned by a magnetic field applied parallel to the film surfaces.

\vspace{1pc}
\end{abstract}


%#####################################################################3
\maketitle

\section{Introduction}
\label{sec.int} 

Possible applications of superconductivity to electronic devices based on the control of vortex motion has received a great deal of attention lately. One line of work uses  vortex pinning by periodic potentials lacking spatial inversion symmetry to produce drift vortex motion in a prefered direction when driven by a oscillating  force with zero average. This is the so called ratchet effect \cite{rev1}. Applications to  removal of trapped flux from superconductors, and  to voltage rectification  have been proposed \cite{tp1,tp2,tp3,tp4,tp5,tp6}. Experimental realizations of the ratchet effect for vortices have been reported by some groups \cite{ep1,ep2,ep3}. The work so far has concentrated on  vortex pinning potentials resulting from periodic modulations of the film thickness or from periodic arrays of anti-dots or blind holes with asymmetric shape. 
The objective of this paper is to show that the  ratchet effect can be obtained for vortices  in thin superconducting films pinned by periodic arrays of magnetic dots  placed in the vicinity of the film.  Here a simple model is studied in detail. One vortex in a thin superconducting film interacting with a periodic array of point magnetic dipoles, placed on top of the film in the London limit. The ratchet effect results from the interaction between the vortex and the magnetic dipole array if the magnetic moments are parallel to the film surfaces, and  all moments point in the same direction. As shown here, the interaction between the vortex and the dipole array breaks inversion symmetry, and  depends on the orientation of the magnetic moment   with respect to the lattice. The latter allows the  ratchet effect to be tuned by rotating the magnetic moments. If the dipole array is made of freely rotating magnetic moments,  the ratchet effect can be produced and tuned by a magnetic field applied parallel to the film surfaces. This field  orients all magnetic moments in the same direction, and  does not influence the vortices  because the film is thin.  This may be of practical interest  because it is possible to fabricate arrays   with freely rotating magnetic moments, as demonstrated recently  by Cowburn, et. al.\cite{ckaw}. These authors reported on the magnetic properties of arrays of nanomagnets made of Supermalloy,  each nanomagnet being a thin circular disk of radius $R$, and found that for $R\sim 50-100$nm the magnetic state of each nanomagnet is a single domain one, with the magnetization parallel to the disk plane, and that the magnetization can be reoriented by small applied fields. They concluded that each nanomagnet acts like a giant magnetic moment free to  rotate. 

The calculations carried out in this paper start from the exact interaction potential in the London limit between a vortex in a thin superconducting film and the dipole array.  The interaction potential depends on the orientation of the magnetic moments  with respect to the dipole lattice and, in general, lacks spatial inversion symmetry.   
The equation of motion for the vortex driven by a force periodic in time is then solved numerically at zero temperature, both for  sinus-wave and square-wave time dependencies. It is found that drift vortex motion takes place in such a way that during each period of oscillation the vortex displacement is equal to a translation vector of the dipole array. 
The vortex displacement is  not, in general, in the direction of the driving force, and depends in a complicated way on the orientation of the magnetic moments and of the driving force with respect to the dipole lattice. The ratchet effects is, thus, two-dimensional in general. One-dimensional ratchet effects are obtained if the driving force oscillates in a direction of high symmetry of the dipole array lattice. In this case the vortex motion is found to be similar to that reported in the literature for particles interacting with one-dimensional ratchet potentials \cite{rev1}. The vortex drifts in the direction of  the driving force oscillation. Its displacement during each period of oscillation is a integer multiple of the dipole lattice period in this direction, and depends on the orientation of the magnetic moments with respect to the direction of the driving force.  The paper argues that the results obtained for this simple model  are applicable to thin low-$T_c$ superconducting  films, and can be extended to vortex lattices commensurate with the dipole array.

The paper is organized as follows. In Sec.\ \ref{sec.vdi} the interaction between the vortex and the dipole array is obtained. The motion of the vortex interacting with the dipole array is examined in Sec.\ \ref{sec.rte}, where the main results of the paper are reported. Finally, Sec.\ \ref{sec.dis} discusses  the limits of validity of the model and its possible extensions,  and states the conclusions of the paper.

\section{Vortex- dipole array interaction}
\label{sec.vdi} 

\begin{figure}[t]
\centerline{\includegraphics[scale=0.2]{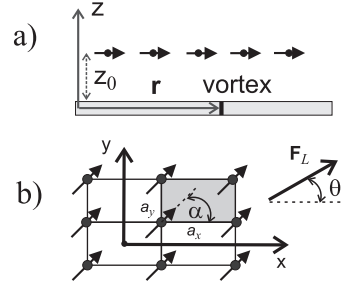}}
\vspace{5mm}
\caption {a) Schematic side view of the superconducting film with one vortex at ${\bf r}$ and the dipole array on top. b) Top view of the dipole array, and definitions of angles $\alpha$ and $\theta$}
\label{fig.fig1}
\end{figure}

The superconductor film is assumed to be planar, with surfaces parallel to each other and to the $x-y$ plane,  isotropic,  characterized by the penetration depth $\lambda$, and of thickness $d\ll \lambda$. A plane dipole array is located above the film at a height $z_0>0$. The dipole array is characterized by a rectangular unit cell, with sides $a_x$ and $a_y$, and  one dipole per cell. All cells have the same magnetic moment, ${\bf m}$, parallel to the film surfaces, and   oriented at an angle $\alpha$ with respect to the $x$-axis, as shown in Fig.\ \ref{fig.fig1}. In the case of freely rotating magnetic moments, a magnetic filed ${\bf H}$ must be applied parallel to the film surfaces, at an angle $\alpha$  with the $x$-axis, in order to orient all magnetic moments in the same direction. The magnitude of the field must be large enough to avoid reorientation of the dipoles by the fields created by the vortex, by the other dipoles, and by the screening and transport currents flowing in the film \cite{gmc2}. The field   ${\bf H}$ does not influence the vortex, because the film is  thin 
($d\ll\lambda$).      
%
%################################################################################# 
%######################################################################
%

%######################## NEW ##################################################
The London limit interactions between vortices in superconducting films and magnetic dipoles located outside the film have been discussed in detail in Ref.\cite{gmc1}, based on the exact solutions of London equations obtained in 
Ref.\cite{gmcehb}. The interactions  result from the action of the screening current induced by the inhomogeneous field generated by the dipoles on the vortices, and can be expressed as the  magnetostatic interaction of the dipoles with the magnetic field generated by vortices. 
According to these results,  the energy of interaction for a  single vortex, with unity vorticity, located at position ${\bf r}$, and the  dipole array,  can be written as 
%######################## end ##################################################
 \begin{equation}  
U_{v-da}({\bf r}) = - \sum_{j=1,N}{\bf m}\cdot{\bf b^s_{\perp}}({\bf R}_j-{\bf r}) 
  \label{eq.uvd}
  \end{equation} 
where $N$ is the number of dipoles in the array, ${\bf R}_j$ their positions and ${\bf b^s_{\perp}}$ is the component parallel to the film surfaces of the field generated by the vortex.  The vortex field is given in the thin film limit and for  $r \ll \Lambda=2\lambda^2/d\;$ by \cite{gmcehb}  
 \begin{equation}  
  {\bf b}^s_{\perp}({\bf r}) =-\frac{\phi_0 d}{4\pi \lambda^2}\, \frac{{\bf r}}{r^2}(1-\frac{z_0}{\sqrt{r^2+z^2_0}})
 \, .
  \label{eq.bvt} 
  \end{equation}
For $r > \Lambda\;$, ${\bf b}^s_{\perp}({\bf r})\sim {\bf r}/r^3$. For a periodic dipole array,  the most important contribution to $U_{v-da}$ comes from dipoles close to the vortex. The interaction between the vortex and a dipole located at distances larger than $\Lambda$ from it behaves as 
$({\bf R}_j-{\bf r})/\mid {\bf R}_j-{\bf r}\mid^3$, thus falling off as $1/R^2_j$ for 
$R_j \gg r$. However, the singular $1/R^2_j$ dependence is canceled out in the sum over dipoles present in Eq.\ (\ref{eq.uvd}) by symmetry. The leading contribution falls only as $1/R^3_j$. Here the contribution from far away dipoles is cutoff exponentially, as discussed shortly. 
%################################################################################# 
\begin{figure}[t]
\centerline{\includegraphics[scale=0.3]{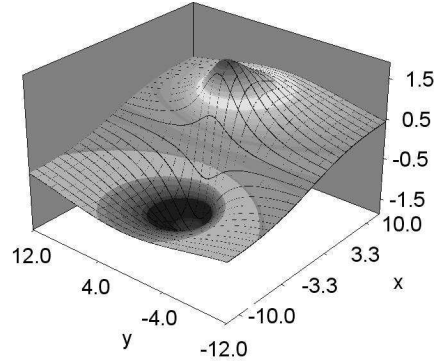}}
\vspace{5mm}
\caption { Vortex pinning potential (in units of  $\epsilon_0 d=(\phi_0/4\pi\lambda)^2d$) for a single dipole with 
${\bf m}$ parallel to the $x$-axis.  $x$ and $y$ in units of $\xi$. Parameters: $m=0.25\phi_0z_0$, $d=z_0=2\xi$, 
$\lambda=10\xi$.}
\label{fig.fig2}
\end{figure}
%######################################################################
%

The vortex pinning potential can be written as the sum of contributions from each dipole, that is 
 \begin{equation}  
U_{v-da}({\bf r}) = - \sum_{j=1,N}{\cal U}({\bf R}_j-{\bf r})\; , 
  \label{eq.uvdb}
  \end{equation}
where ${\cal U}({\bf r})$ is the pinning potential for a single dipole located at the origin, and given by
 \begin{equation}  
{\cal U}({\bf r}) = \frac{\phi_0 d}{4\pi \lambda^2}\, \frac{{\bf m}\cdot{\bf r}}{r^2}(1-\frac{z_0}{\sqrt{r^2+z^2_0}})e^{-(r/\Lambda)} \; ,
  \label{eq.u1d}
  \end{equation}   
where the exponential factor is the cutoff for the contributions from far away dipoles. 
According to Eq.\ (\ref{eq.u1d}), ${\cal U}({\bf r})$ is anti-symmetric with respect to spatial inversion and has a minimum (maximum) located at  ${\bf r}\cdot\hat{\bf m}=-(+)1.3z_0$.  For ${\bf m}$ oriented parallel to the $x$-axis, ${\cal U}({\bf r})$  has the spatial dependence shown in Fig.\ \ref{fig.fig2}. 

\begin{figure}[h]
\centerline{\includegraphics[scale=0.35]{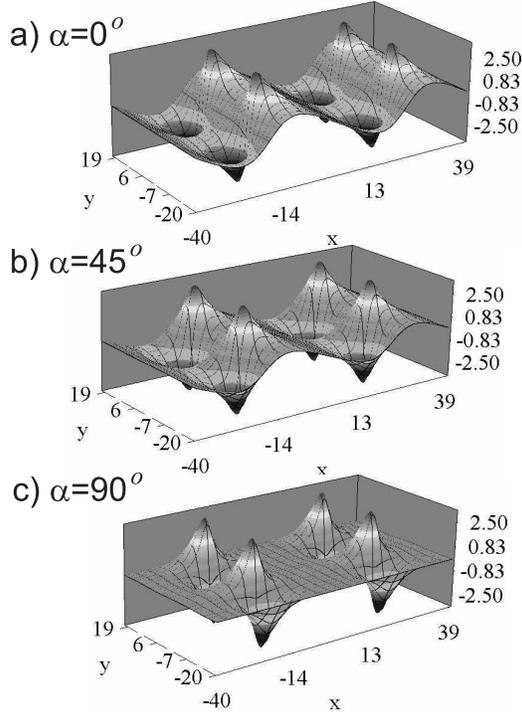}}
\vspace{5mm}
\caption  {Vortex pinning potentials (in units of $\epsilon_0 d$) for  dipole arrays with magnetic moments oriented at  angle $\alpha$ with the $x$-axis.  $x$ and $y$ in units of $\xi=$vortex core radius. Parameters: $m=0.25\phi_0z_0$, $d=z_0=2\xi$, 
$\lambda=10\xi$. } 
\label{fig.fig3}
\end{figure}

For a  dipole array   $U_{v-da}({\bf r})$ is the periodic repetition of ${\cal U}$, centered at each dipole. The spatial dependence of $U_{v-da}({\bf r})$ for a typical rectangular dipole array is shown in Fig.\ \ref{fig.fig3}  for some values of $\alpha$.

The vortex-dipole array interaction $U_{v-da}({\bf r})$ lacks spatial inversion symmetry and leads to the  ratchet effect  when the vortex is driven by an oscillating force, as discussed next.

%
%################################################################################# 
%######################################################################

\section{Ratchet effects}
\label{sec.rte}

The motion of a vortex in the potential $U_{v-da}({\bf r})$, driven by a force ${\bf F}_L(t)$ is governed, at zero temperature, by
\begin{equation}
\eta \frac{d{\bf r}}{dt}= {\bf F}_L(t)  -{\mbox{\boldmath ${\nabla}$}}U_{v-da}({\bf r}) \; ,
\label{eq.lan}  
\end{equation} 
where $\eta$ is the vortex friction coefficient. The driving force, ${\bf F}_L(t)$, results from a transport current density, ${\bf J}(t)$ applied to the film, that is ${\bf F}_L(t)=(\phi_0 d/c){\bf J}(t)\times \hat{{\bf z}}$. The dipole array lattice is assumed rectangular with  $a_x>a_y\gg z_0$. In this case $U_{v-da}({\bf r})$ is like that shown in 
Fig.\ \ref{fig.fig3}. It is convenient to use  the following natural units for physical quantities. 
Length: $\xi=$ vortex core radius. Force:  $\epsilon_0=(\phi_0/4\pi\lambda)^2$. Time:  $\tau=\eta\xi/\epsilon_0$. Velocity: $\epsilon_0/\eta$. Current density: $J_d=c\phi_0/(12\sqrt{3}\pi^2\lambda^2\xi)$, where  $J_d$ is the depairing current.
When  Eq.\ (\ref{eq.lan}) is written in terms of these natural units it depends only on the scaled parameters 
$m/\phi_0z_0$, $d/z_0$, $\Lambda/z_0$, $J/J_d$, and  on $ ({\bf R}_j-{\bf r})/z_0$. The values of $J$ are, of course, limited to $J<J_d$, but in the results reported next regions where $J_c>J_d$ are  discussed for the sake of completeness.

\begin{figure}[h]
\centerline{\includegraphics[scale=0.2]{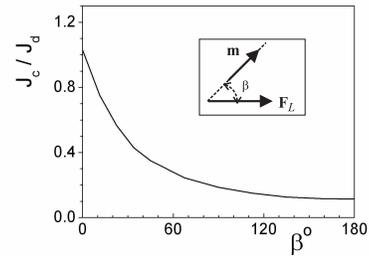}}
\vspace{5mm}
\caption  {Critical current, $J_c$, for a vortex pinned by the dipole array vs.  angle  between the magnetic  moments and the driving force, $\beta$ (inset).} 
\label{fig.fig4}
\end{figure}

As will be seen shortly, the dc critical current to depin a vortex from a  minimum of $U_{v-da}({\bf r})$ plays an important role in the ratchet effect. To obtain it, 
Eq.\ (\ref{eq.lan}) is solved for a time-independent transport current. The critical current  depends, in general, on the orientations of ${\bf m}$ and of the driving force with respect to the dipole lattice. In the case of interest here, $a_x,\,a_y\gg z_0$, it is found that the critical current is essentially identical to that for a vortex pinned by a single dipole, obtained in Ref. \cite{gmc2}, and depends only on the angle between the dipole and the driving force, 
$\beta$. The critical current $J_c(\beta)$ is shown in Fig.\ \ref{fig.fig4}. 
The maximum  $J_c$ occurs when ${\bf m}$ is parallel to the direction of the driving force, $\beta=0$, and the  minimum  for $\beta=180^o$. There is  a  tenfold difference between the maximum and minimum values, and a smooth decrease in $J_c$ with increasing $\beta$. The maximum and minimum $J_c$ can be estimated analytically from the single dipole pinning potential, Eq.\ (\ref{eq.u1d}). The result is  $J_c/J_d\simeq 4m/\phi_0z_0$   for $\beta=0$, and $J_c/J_d\simeq 0.4m/\phi_0z_0$ for  $\beta=180^o$.
%
%################################################################################# 
%######################################################################

For an oscillating transport current with zero average, Eq.\ (\ref{eq.lan})  is solved numerically and the vortex velocity  averaged in time in the steady-state regime, denoted by ${\bf V}$, is calculated. Drift vortex motion occurs if ${\bf V}\neq 0$. The transport current is assumed to oscillate in a direction defined by the unit vector $\hat{{\bf n}}$, that makes an angle $\theta$ with the $y$-axis, and to be periodic in time   with period $P$ and zero average.
Writing  ${\bf J}(t)=J_T f(t)\hat{{\bf n}}$, $J_T$ is the amplitude, and $f(t)$ is chosen as either as sinus wave, $f(t)=\sin{(2\pi t/P)}$, or as square wave, $f(t)=1$ for $kP\leq t <(k+1/2)P $, and $f(t)=-1$ for $ (k-1/2)P\leq t <kP $, with $k=0,\pm 1,\pm 2,...$.
The driving force oscillates in the direction perpendicular to $\hat{{\bf n}}$, which  makes an angle  $\theta$ with the $x$-axis (Fig.\ \ref{fig.fig1}). It is found that for arbitrary $\theta$ the ratchet effect is two-dimensional (2D), with drift vortex motion taking  place both in the directions parallel and perpendicular to ${\bf F}_L(t)$.  The ratchet effect is one-dimensional (1D) only if  $\theta=0^o,\; 90^o$ , that is, when  ${\bf F}_L(t)$ oscillates along directions of high symmetry of the rectangular dipole lattice ($x$ and $y$ directions, respectively). In these cases drift vortex motion takes place only in the direction of the driving force. 

Next the results of the numerical solution of  Eq.\ (\ref{eq.lan}) are reported for the following parameter values: 
 $d=z_0=2\xi$, $\lambda=10\xi$, $a_x=20\xi$, $a_y=40\xi$, $m=0.25\phi_0z_0$.

\subsection{One-dimensional ratchet}
\label{sec.1dr} 

\begin{figure}[t]
\centerline{\includegraphics[scale=0.3]{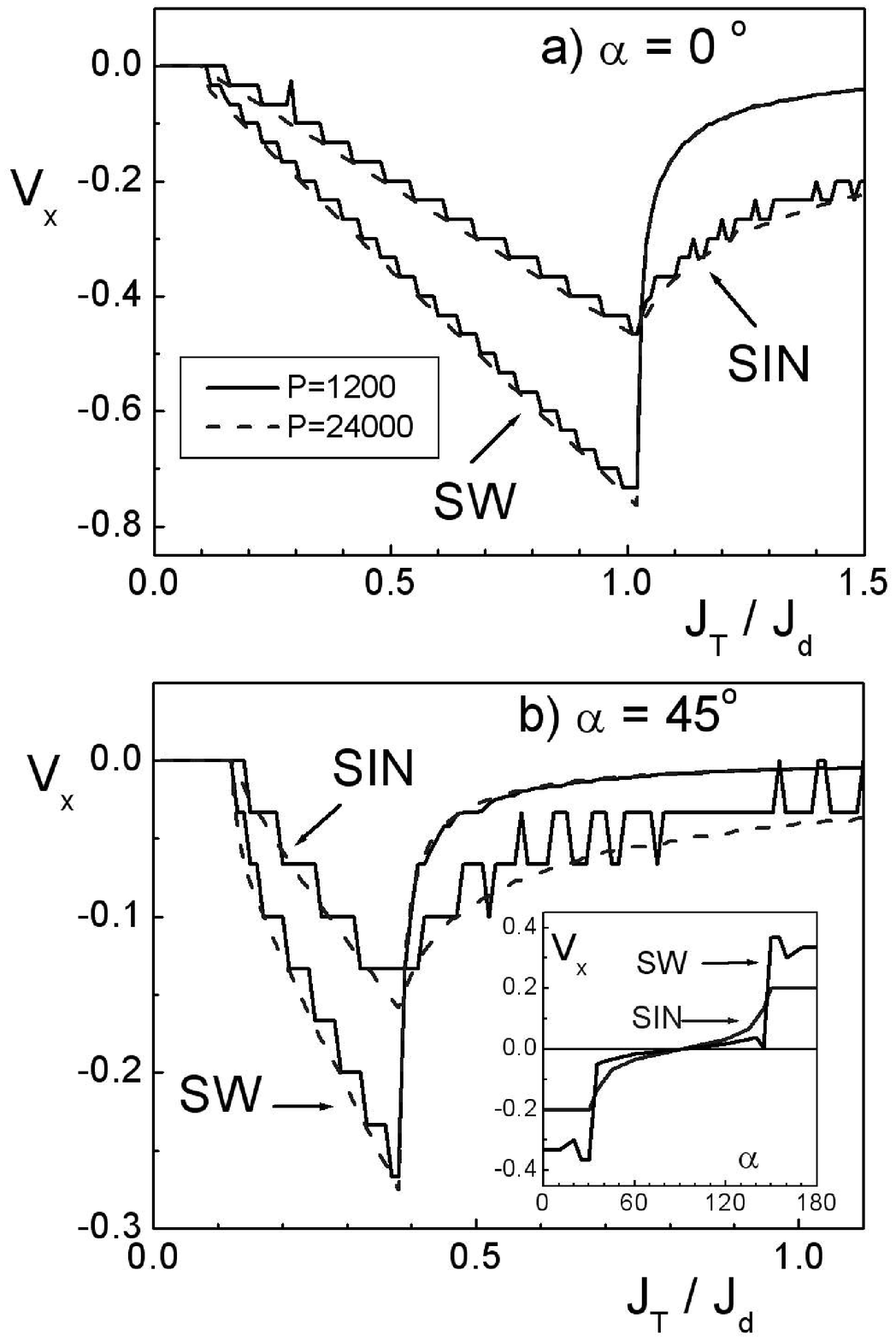}}
\vspace{5mm}
\caption  {Vortex drift velocity (in units of $\epsilon_0/\eta$) vs. $J_T$ for driving force  parallel to the $x$-axis ($\theta=0$), and  magnetic moments at an angle $\alpha$ with the $x$-axis. Labels: SIN and SW = sinus-wave and  square-wave driving forces, respectively. Inset in b) $V_x$ vs. $\alpha$ for $J_T=0.5J_d$ and $P=1200$.  $P$ in units of $\tau=\eta\xi/\epsilon_0$.  } 
\label{fig.fig5}
\end{figure}

Here the 1D ratchet effect is considered for  $\theta=0$, that is ${\bf F}_L(t)$ oscillating in the $x-$direction. In this case the vortex drifts only in the $x-$direction ( $V_y=0$), and  $V_x$ depends on  $J_T$,  $\alpha$ and $P$. Typical $V_x$ vs.  $J_T$ curves  for $0\leq\alpha<90^o$ are shown in Fig.\ \ref{fig.fig5}. In this case the vortex drifts in the negative $x$-direction. For $90^o<\alpha\leq180^o$, the vortex drifts in the positive $x$-direction, and the  $V_x$ vs. $J_T$ curves are identical to those  for $180^o-\alpha$  with  $V_x$ replaced by $-V_x$. For $\alpha=90^o$ there is no drift vortex motion, and $V_x=0$. The most important characteristics of the $V_x$ vs.  $J_T$ curves  for $0\leq\alpha<90^o$ are the following. Drift vortex motion only occurs if $J_T$ is larger than a minimum value, denoted by $J_0(\alpha)$. For $J_T>J_0(\alpha)$ the  $V_x$ vs.  $J_T$ curves have two distinct regions:  one for $J_0(\alpha)<J_T<J_m(\alpha)$, and another for $J_T>J_m(\alpha)$, where $J_m(\alpha)$ denotes  the value of $J_T$ for which $V_x $ is minimum ($\mid V_x \mid $ is maximum). It is found that in the region $J_0(\alpha)<J_T<J_m(\alpha)$ 
the vortex displacement during each period is an integer multiple of the lattice parameter $a_x$. The  $V_x$ vs.  $J_T$ curve consists of a series of plateaux where $V_x$ is independent of  $J_T$. In each plateau $V_x=\ell a_x/P, \ell=1,2,3,...$. As shown in Fig.\ \ref{fig.fig5}, the plateaus are clearly visible in the curve for 
$P=1200 \tau$. For $P=24000 \tau$ the plateaux are too narrow to  be distinguished in the scale of the figure.
For $J_T>J_m(\alpha)$, the $V_x$ vs. $J_T$ curve  is strongly affected by the time dependence of the driving force. For the square wave, $V_x$ drops to zero very quickly with $J_T$, while for the sinus wave  $V_x$ drops to zero slowly and is a complicated function of  $J_T$. The dependence of $V_x$ on $\alpha$ for fixed $J_T$ is shown in the inset of 
Fig.\ \ref{fig.fig5}.b, and will be discussed in more detail later.
It is found that for both time dependencies $J_0(\alpha)\simeq J_c(180^o-\alpha)$ and $J_m(\alpha)\simeq J_c(\alpha)$. More details of the  dependence of $V_x$  on  $J_T$,  $\alpha$ and $P$ is given next.

%
%################################################################################# 
%######################################################################
%
%################################################################################# 
\begin{figure}[t]
\centerline{\includegraphics[scale=0.2]{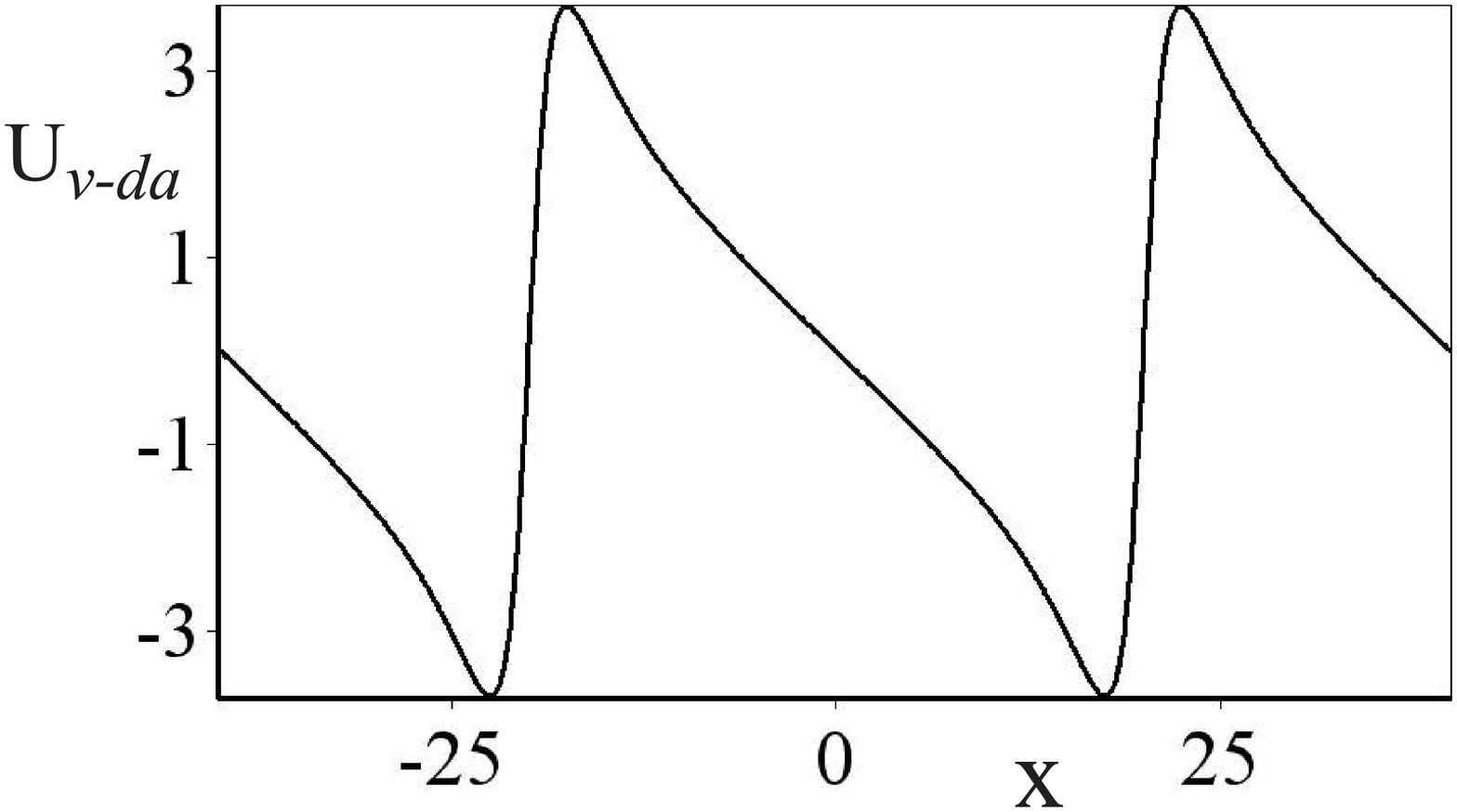}}
\vspace{5mm}
\caption  {One-dimensional vortex pinning potential $U_{v-da}(x,y=a_y/2)$ (in units of $\epsilon_0d$), for 
$\alpha=0$. $x$ in units of $\xi$} 
\label{fig.fig6}
\end{figure}
%######################################################################
%################################################################################# 

$\alpha=0$: In the steady state regime there is no motion in the $y-$ direction. The vortex moves only in negative $x$-direction,  along a line  connecting the minima of  $U_{v-da}({\bf r})$, for instance $y=a_y/2$ (see Fig.\ \ref{fig.fig3}.a). Thus the vortex dynamics is one-dimensional in the potential  $U_{v-da}(x,y=a_y/2)$ for $\alpha=0$. As shown in Fig.\ \ref{fig.fig6}, this 1D potential has  characteristics similar to ratchet potentials used in other contexts \cite{rev1}. For $J_0(0)<J_T<J_m(0)$ the vortex moves only during the half periods of oscillation when ${\bf F}_L(t)$ points in the negative $x$-direction. When ${\bf F}_L(t)$ points in the positive $x$-direction the vortex is pinned because 
$J_m(0)\simeq J_c(0)$ (see Fig.\ \ref{fig.fig4}). For $J_T>J_m(0)$ the vortex moves both in the negative and positive $x-$directions, but the displacement is larger in the negative $x-$direction due to the lack of inversion symmetry of $U_{v-da}(x,y=a_y/2)$. 

\noindent  $0<\alpha<90^o$: Vortex motion is found to be  two-dimensional, but drift takes  place only along the negative $x-$direction. In the $y$-direction the vortex oscillates with zero average. Drift vortex motion is similar 
to that for $\alpha=0$. For $J_0(\alpha)<J_T<J_m(\alpha)$ the vortex is pinned when ${\bf F}_L(t)$ points in the positive $x$-direction because $J_m(\alpha)\simeq J_c(\alpha)$. However, since $J_c(\alpha)$ decreases as $\alpha$ increases, the value of $J_T$ for which  $V_x$ is minimum ( $J_T=J_m(\alpha)$ ) decreases with $\alpha$,  and  the value of $V_x$ at the minimum increases ($\mid V_x\mid $ decreases) with $\alpha$. As shown in the inset of  Fig.\ \ref{fig.fig5}.b, for $J_T$ fixed, the value of $V_x$ is weakly dependent on $\alpha$ as long as $J_T<J_m(\alpha)$. 
For $J_T>J_m(\alpha)$,  $V_x$ drops to zero rapidly with $\alpha$. This result can be understood by noting first that the average slopes of the $V_x$ vs. $J_T$ curves  in region $J_0(\alpha)<J_T<J_m(\alpha)$ are essentially independent of $P$, as shown in Fig.\ \ref{fig.fig5}. Second, the average slope of the $ V_x$ vs. $J_T$ curve is linear, except for $J_T$ close to $J_0(\alpha)$, and weakly dependent on $\alpha$. This occur because the  driving force is large compared with the pinning force in the direction of drift. An estimate for $V_x$ in this case is obtained by neglecting the pinning force during the half period of oscillation where  vortex drift motion takes place. The result is  $ V_x\sim -0.5 J_T/J_d(\epsilon_0 /\eta)$ for the sinus wave,  and 
$ V_x\sim -0.8 J_T/J_d(\epsilon_0 /\eta)$ for the square wave. These estimates are in reasonable agreement with the values of $V_x$ shown in the inset of Fig.\ \ref{fig.fig5}.b for $J_T<J_m(\alpha)$

%
%################################################################################# 
\begin{figure}[t]
\centerline{\includegraphics[scale=0.3]{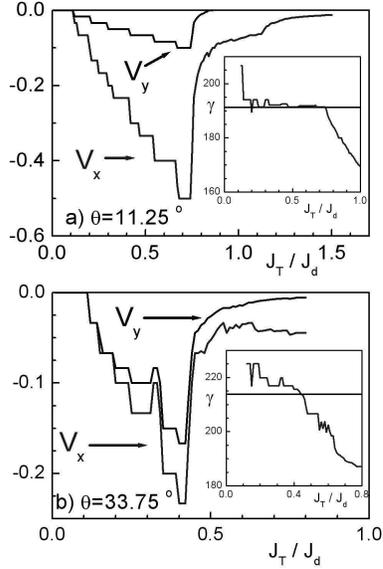}}
\vspace{5mm}
\caption  {Vortex drift velocity components (in units of $\epsilon_0/\eta$) vs. $J_T$ for magnetic moments parallel  to  the $x$-axis ($\alpha=0$), and square-wave driving force at angle $\theta$ with the $x$-axis and period 
$P=1200\tau$. Insets: angle  between the vortex drift velocity and $x$-axis,  $\gamma$, vs. $J_T$. Horizontal lines: $\gamma=\theta+180^o$.  } 
\label{fig.fig7}
\end{figure}
%######################################################################
 
%
%################################################################################# 
\begin{figure}[t]
\centerline{\includegraphics[scale=0.3]{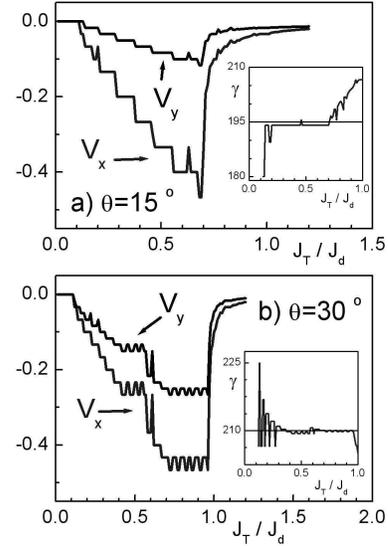}}
\vspace{5mm}
\caption  {Vortex drift velocity components (in units of $\epsilon_0/\eta$) vs. $J_T$ for  magnetic moments  at  $\alpha=30^o$ with the $x$-axis, and  square-wave driving force at angle $\theta$ with the $x$-axis and  period $P=1200\tau$. Insets: angle  between the vortex drift velocity and $x$-axis,  $\gamma$, vs. $J_T$. Horizontal lines: $\gamma=\theta+180^o$. } 
\label{fig.fig8}
\end{figure}
%######################################################################

Tuning  the 1D ratchet effect is illustrated in the inset of Fig.\ \ref{fig.fig5}.b.  
Basically, by rotating the magnetic moments the vortex drift velocity can be switched  off, on, and reversed.

\subsection{Two-dimensional ratchet}
\label{sec.2dr} 

Now the 2D ratchet effect for $\theta\neq 0$ is considered. In this case ${\bf V}$ depends both on $\theta$ and 
$\alpha$, as well as on $J_T$, in a complicated way. Typical results for the dependence of  ${\bf V}$ on  $J_T$ and $\theta$ is shown in Fig.\ \ref{fig.fig7} and Fig.\ \ref{fig.fig8}  for the square-wave driving force. The curves for 
$V_x$ and $V_y$ vs. $J_T$ are similar to those obtained  for  $\theta=0$. There is a value  $J_T=J_0$ below which $V_x=V_y=0$, and a value $J_T=J_m$ for which both $V_x$ and $V_y$ are minimum. It is found that   $J_m\simeq J_c(\mid \theta -\alpha\mid )$, and $J_0\simeq J_c(180^o-\mid \theta -\alpha\mid)$.
For $J_0<J_T<J_m$, the $V_x$ vs. $J_T$ and $V_y$ vs. $J_T$ curves  show plateaux where both  $V_x$ and $V_y$ are independent $J_T$. Each plateaux correspond to a vortex displacement by a lattice translation vector, ${\bf R}=n_x a_x \hat{{\bf x}} + n_y a_y \hat{{\bf y}},\; n_x,\,n_y= 0,\pm 1,\pm 2,...$. 
The direction of ${\bf V}$ (angle $\gamma$ with the $x-$axis) is not, in general, parallel to the direction of  driving force oscillation, as shown in the insets in  Fig.\ \ref{fig.fig7} and Fig.\ \ref{fig.fig8}. However, as shown in Fig.\ \ref{fig.fig7}.a ( $\alpha=0,\, \theta=11.25^o$) and Fig.\ \ref{fig.fig8}.a. ($\alpha=30^o,\, \theta=15^o$), 
 there are regions where the two directions nearly coincide ($\gamma=180^o+\theta$).  This occurs because for $J_T$ in these regions the driving force is large compared to the pinning force during the drift vortex motion.  Similarly to the case $\theta=0$ discussed above, the average slopes of the $V_x$ and $V_y$ vs. $J_T$ curves are linear in these regions.   For $\alpha=0,\, \theta=33.75^o$, $\gamma$  differs considerably from that of driving force oscillation. For $\alpha=30^o,\, \theta=30^o$, there is also a region where  ${\bf V}$ is parallel to the direction of drive, but the average slopes of the  $V_x$ vs. $J_T$ and $V_y$ vs. $J_T$ curves are not linear. In this case the driving force oscillates in a direction parallel to that of the magnetic moments, but the dependence of the drift velocity on $J_T$ differs considerably from that for $\alpha=\theta=0$. This  shows that ${\bf V}$ depends both on $\theta$ and $\alpha$.  Tuning of the  vortex drift velocity by rotating the direction of the magnetic moments is also possible for 
$\theta \neq 0$. The dependence of $V_x$ on $\alpha$ for fixed $J_T$ is, essentially, similar to that shown in  
the inset of Fig.\ \ref{fig.fig5}.b, since the $V_x$ vs. $J_T$ curves for $\theta\neq 0$ have the same basic structure as that for $\theta=0$.

\section{Discussion}
\label{sec.dis} 

First, an order of magnitude estimate of the vortex drift velocities obtained above is given. Consider the case $\theta=0$ and assume that $\xi=20nm$, and $\eta=\eta_0d$, with 
$\eta_0=7\times 10^{-6}\, Nsm^{-2}$ \cite{tp1}. In this case  $d=z_0=2\xi=40nm$, $\lambda=10\xi=200nm$ and $m=0.25\phi_0z_0\sim 10^7\mu_B$ ($\mu_B=$Bohr magneton). According to Sec.\ \ref{sec.rte}, the  maximum values of  $\mid V_x\mid$ are $\mid V_x\mid_{max} \sim 0.5J_c(\alpha)/J_d(\epsilon_0/\eta)\sim 10J_c(\alpha)/J_d\, m/s$
for the sinus wave and $\mid V_x\mid_{max} \sim 0.8J_c(\alpha)/J_d(\epsilon_0/\eta)\sim 16 J_c(\alpha)/J_d\, m/s$ for the square wave. This gives for the sinus wave $\mid V_x\mid_{max}\sim 10 m/s$ for $\alpha=0$, and $\mid V_x\mid_{max}\sim 4 m/s$ for $\alpha=45^o$, and $\mid V_x\mid_{max}\sim 16 m/s$ for $\alpha=0$, and $\mid V_x\mid_{max}\sim 6 m/s$ for $\alpha=45^o$, for the square wave. These velocity  values  are of the same order of magnitude as those reported in Ref. \cite{tp1}.  The value of the unit time is $\tau=\eta\xi/\epsilon_0\sim 10^{-9}s$. Thus, the transport current oscillation
frequencies, $\nu=1/P$, used in the calculations are $\nu=800\,KHz,\; 40\,KHz $ for $P=1200\tau,\; 24000\tau$, respectively.    

The results obtained in this paper are believed to be representative of low-$T_c$ superconducting films with magnetic dipole arrays placed on top. First, the particular set of  parameters  used, $d\sim z_0\sim \xi$, are typical ones.  For instance, in the experiments with arrays of magnetic dots with permanent magnetization placed on top of superconducting Nb films, reported in Ref.\cite{pann1},  $d=20nm\sim\xi$. The  magnetic dots  are separated from the film by a thin protective layer of thickness $\sim 20nm$, so that the distance from the magnetic dipole to the film is $z_0\sim \xi$.  Second, since the vortex drift velocity depends on the scaled parameters $m/\phi_0z_0$, $d/z_0$, $J/J_d$, 
many superconducting film-dipole array systems are equivalent. Third, the results obtained in this paper do not depend on the particular dipole array used.  What is essential for the ratchet effect is the periodic vortex pinning potential with broken spatial inversion symmetry created by it. Most dipole arrays will do, since the interaction between the vortex and a single dipole lacks inversion symmetry. 

The London limit is valid for vortices in low-$T_c$ films. However, when a magnetic dipole is placed close to the film, it certainly breaks down if the dipole field  destroys superconductivity locally in the film. Roughly speaking, London theory is valid as long as the maximum dipole field at the film is less than the upper critical field, that is, $m/z^3_0 < \phi_0/(2\pi \xi^2)$, or $m/(\phi_0z_0) < (z_0/\xi)^2/2\pi$. For the parameters used in the above calculations ($z_0=2 \xi$) this gives $m/(\phi_0z_0) < 0.64$, which is larger than the value of $m$  used in this paper. The London limit would be a  better approximation if the present calculations were carried out for larger values of $z_0/\xi$. However, the  results  would be identical to those described above if $m$ and $d$ were scaled by the same factor as  $z_0/\xi$. For instance, if  $z_0\rightarrow 2z_0$, $J_c/J_d$ would remain the same if  $d\rightarrow 2d$ and $m\rightarrow 2m$, but the upper limit of  $m/\phi_0z_0$ for the  validity of the London approximation would increase by a factor of $4$. The present model also breaks down if $m$ is sufficiently large to create vortices in the film. The threshold value of $m$ for spontaneous vortex creation,  estimated as $m\sim 0.7\phi_0z_0$ using the results of Ref.\cite{gmc1}, is larger than $m$ used here.  

The results obtained in this paper for a single vortex  also apply to vortex lattices pinned by dipole arrays if the vortex lattice is commensurate with the dipole array and if the vortex density is at most one vortex per dipole. In this case the driving force is expected to move the vortex lattice as a whole, that is preserving the spatial order, so that  the effects of vortex-vortex interactions are negligible.

In conclusion then, this paper demonstrates the ratchet effect for  vortices in a thin superconducting film pinned by  periodic arrays of magnetic dipoles placed on top of the film, and shows that the ratchet effect can be tuned by rotating the magnetic moments. In the case of magnetic moments free to rotate, the ratchet is created and tuned  by a magnetic film applied parallel to the film surfaces.


\begin{thebibliography}{9}

\bibitem{rev1} For a recent review, see P. Reimann Phys. Rep. {\bf 361}, 57 (2002).

\bibitem{tp1} C.S.Lee, B.Jank{\'o}, L.Der{\'e}nyi, A.L.Bar{\'a}basi, Nature {\bf 400} , 337 (1999).


\bibitem{tp2} I.Zapata, R.Bartussek, F.Sols, P.H$\ddot{\rm a}$nggi, Phys. Rev. Lett. {\bf 77}, 2292 (1996).

\bibitem{tp3} J.F.Wambaugh, C.Reichhardt, C.J.Olson, F. Marchesoni, F.Nori, Phys. Rev. Lett. {\bf 83}, 5106 (1999).

\bibitem{tp4} B.Y.Zhu, F.Marchesoni, V.V.Moshchalkov, F.Nori, Phys.Rev.B {\bf 68} , 014514 (2003).

\bibitem{tp5} B.Y.Zhu, F.Marchesoni, F.Nori, Physica E {\bf 18}, 318 (2003).

\bibitem{tp6} S.Savel$^{\prime}$ev, F.Nori, Nature Mater. {\bf 1}, 179 (2003).

\bibitem{ep1} J.E. Villegas, S.Savel$^{\prime}$ev, F. Nori, E.M. Gonzalez, J.V. Anguita, R. Garc{\'{\i}}a , J.L. Vicent,
Science {\bf 302}, 1188 (2003).

\bibitem{ep2} J.E. Villegas,  E.M. Gonzalez, M. P. Gonzalez, J.V. Anguita, J.L. Vicent,
Phys.  Rev. B {\bf 71}, 024519 (2005).

\bibitem{ep3} J. Van de Vondel, C. C. de Souza Silva,  B. Y. Zhu,  M. Morelle,  and V. V. Moshchalkov, 
Phys. Rev. Lett. {\bf 94}, 057003 (2005).

\bibitem{ckaw} R.P Cowburn, D.K. Kolstov, A.O. Adeyeye, and M.E. Welland, Phys. Rev. Lett. {\bf 83}, 1042 (1999).
 
\bibitem{gmc1} G. Carneiro, Phys. Rev. B, {\bf 69} 214504 (2004). 

\bibitem{gmcehb} G. Carneiro, and E.H. Brandt, Phys. Rev. B {\bf 61} 6370 (2000). 

\bibitem{gmc2} G. Carneiro, to appear in Europhys. Lett. .  

\bibitem{pann1} Y. Otani, B. Pannetier, J. P. Nozi\`eres,  and D. Givord 
J. Magn. Magn.Mater.{\bf 126}, 622 (1993); 
 Y. Otani, Y. Nozaki, H. Miyajima, B. Pannetier, M. Ghidini, J. P. Nozi\`eres,
G. Fillion, and P. Pugnat, 
Physica C {\bf 235 - 240}, 2945 (1994); 
 Y. Nozaki, Y. Otani, K. Runge, H. Miyajima, B. Pannetier, J. P. Nozi\`eres, and
G. Fillion, J. Appl. Phys{\bf 79}, 11 (1996). 


\end{thebibliography}
\end{document}